\documentclass[twoside,twocolumn,9pt]{article}
\usepackage{extsizes}
\usepackage[super,sort&compress,comma]{natbib}
\usepackage[version=3]{mhchem}
\usepackage[left=1.5cm, right=1.5cm, top=1.785cm, bottom=2.0cm]{geometry}
\usepackage{mathptmx}
\usepackage{sectsty}
\usepackage{graphicx}
\usepackage{lastpage}
\usepackage[format=plain,justification=justified,singlelinecheck=false,font={stretch=1.125,small,sf},labelfont=bf,labelsep=space]{caption}
\usepackage{float}
\usepackage{fancyhdr}
\usepackage{fnpos}
\usepackage[UKenglish]{babel}
\addto{\captionsenglish}{%
  
}
\usepackage{array}
\usepackage{droidsans}
\usepackage{charter}
\usepackage[T1]{fontenc}
\usepackage[usenames,dvipsnames]{xcolor}
\usepackage{setspace}
\usepackage[compact]{titlesec}
\usepackage{hyperref}
\definecolor{cream}{RGB}{222,217,201}
\hypersetup{hidelinks}
\usepackage{authblk} 
\usepackage[utf8]{inputenc}
\usepackage[T1]{fontenc}
    
\usepackage{amsmath,amsthm,amssymb,mathtools}
\usepackage[bbgreekl]{mathbbol}

\usepackage[]{graphicx}
\DeclareGraphicsExtensions{%
    .pdf,.PDF,%
    .png,.PNG,%
    .jpg,.jpeg}
\newcommand{\Cabs}{\sigma_\text{abs}}

\newcommand{\Csca}{\sigma_\text{sca}}
\newcommand{\Cext}{\sigma_\text{ext}}

\newcommand{\vecEsca}{\mathbf{E}^\text{\sc sca}}

\newcommand{\vecEinc}{\mathbf{E}^\text{\sc inc}} 

\newcommand{\vecEsph}{\mathbf{E}^\text{\sc sph}} 
\newcommand{\vecEexc}{\mathbf{E}^\text{\sc exc}} 
\newcommand{\vecEdips}{\mathbf{E}^\text{\sc dip}} 

\newcommand{\vecrgM}{\mathbf{M}^{(1)}} 
\newcommand{\vecrgN}{\mathbf{N}^{(1)}} 
\newcommand{\vecM}{\mathbf{M}^{(3)}} 
\newcommand{\vecN}{\mathbf{N}^{(3)}} 

\newcommand{\vecCmn}{\mathbf{C}_{mn}} 
\newcommand{\vecDmn}{\mathbf{D}_{mn}} 

\newcommand{\vecE}{\mathbf{E}}

\newcommand{\vecp}{\mathbf{p}}

\newcommand{\vecr}{\mathbf{r}}

\newcommand{\expikr}{e^{ik_1r}}

\newcommand{\veceh}{ \mathbf{\hat e}}
\newcommand{\vecrh}{ \mathbf{\hat r}}
\newcommand{\vecph}{ \mathbf{\hat p}}

\newcommand{\epsnot}{ \varepsilon_0}
\newcommand{\eps}{ \varepsilon_1}

\newcommand{\opA}{ \mathbb{A}}
\newcommand{\opG}{ \mathbb{G}}

\newcommand{\opS}{ \mathbb{S}}
\newcommand{\opI}{ \mathbb{I}}

\renewcommand{\opA}{ \bar{\bar{A}}}
\renewcommand{\opG}{ \bar{\bar{G}}}

\renewcommand{\opS}{ \bar{\bar{S}}}
\renewcommand{\opI}{ \bar{\bar{I}}}

\renewcommand{\bbalpha}{ \bar{\bar{\alpha}}}

\newcommand{\ii}{ \mathrm{i}}

\newcommand{\Rsat}{R_\mathrm{sat}}
\newcommand{\Rcore}{R_\mathrm{core}}

\usepackage{float}
\graphicspath{{figures/}}

\title{\Large \bfseries Generalised coupled-dipole model for core-satellite nanostructures}

\author[1,2]{Stefania Glukhova}
\author[1,2]{Eric C. Le Ru}
\author[1,2,3,*]{Baptiste Augui\'{e}}
\affil[1]{School of Chemical and Physical Sciences, Victoria University of Wellington, PO Box 600, Wellington, New Zealand}
\affil[2]{The MacDiarmid Institute for Advanced Materials and Nanotechnology}
\affil[3]{The Dodd-Walls Centre for Photonic and Quantum Technologies}
\affil[*]{E-mail: \texttt{baptiste.auguie@vuw.ac.nz}}

\date{\today}
\begin{document}

\twocolumn[
  \begin{@twocolumnfalse}
    \maketitle

    \begin{abstract}Plasmonic core-satellite nanostructures have recently attracted interest in photocatalytic applications. The core plasmonic nanoparticle acts like an antenna, funnelling incident light into the near-field region, where it excites the smaller satellite nanoparticles with resonantly enhanced absorption. Computer simulations of the optical absorption by such structures can prove challenging, even with state-of-the-art numerical methods, due to the large difference in size between core and satellite particles. We present a generalised coupled-dipole model that enables efficient computations of light absorption in such nanostructures, including those with many satellites. The method accurately predicts the local absorption in each satellite despite being two orders of magnitude weaker than the absorption in the core particle. We assess the range of applicability of this model by comparing the results against the superposition $T$-matrix method, a rigorous solution of Maxwell's equations that is much more resource-intensive and becomes impractical as the number of satellite particles increases.
    \end{abstract}
    \vskip2em

  \end{@twocolumnfalse}
]
\section{Introduction}
\label{sec:introduction}
The thriving field of nanoscience has enriched the study of light-matter interactions with novel metamaterials displaying unique optical properties. The combination of subwavelength particles with different shapes and materials enables the design of hybrid nanostructures with tailored properties, where each component contributes to a synergistic response arising from their combination. Among such structures, core-satellite clusters, also known as planet-satellite or "raspberry" nanostructures, are formed with a core nanoparticle surrounded by  smaller satellite nanoparticles\cite{holler_protein-assisted_2016}. Although the shape of the nanoparticles in such structures need not be limited to spheres\cite{kuttner_sers_2019,ameri_engineering_2023}, it is the more common architecture, as described in many works\cite{tian_precise_2019,rossner_planet-satellite_2020,peng_gold-planetsilver-satellite_2016,pazos-perez_modular_2019,mai_simple_2022,li_facile_2022,holler_protein-assisted_2016}. Core-satellite combinations can vary in their configurations and materials, which makes them especially versatile in multiple applications. For example, polymer core-satellite particles are widely used in super-hydrophobic materials \cite{mai_simple_2022,li_facile_2022,hu_synthesis_2023}. In optical applications, plasmonic particles of noble metals such as Au and Ag hold particular interest as they exhibit localised surface plasmon resonances associated with large optical cross-sections and local field enhancements\cite{le_ru_principles_2008}. Core-satellite nanostructures composed of plasmonic materials have therefore received considerable attention\cite{choi_coresatellites_2012,yoon_surface_2013,holler_protein-assisted_2016,herran_tailoring_2022}, with potential applications including surface-enhanced Raman spectroscopy and sensing\cite{san_juan_synthesis_2022,pazos-perez_modular_2019,gu_core-satellite_2022,beulze_robust_2017}, or optical magnetism \cite{ponsinet_resonant_2015,li_controlling_2018}. More recently, the combination of noble metals with catalytic materials in core-satellite structures has been put forward in the design of antenna-reactor photo-catalysts \cite{Camargo:2021aa}, where the core particle harnesses incident (sun)light and transfers the energy via hot carriers to the catalyst surfaces\cite{tzounis_temperature-controlled_2019,tian_precise_2019,li_revisiting_2013,herran_tailoring_2022,filie_dynamic_2021,bu_hydrogen_2019}.

Core-satellite structures are inherently multi-scale, and the overall optical response is typically dominated by the larger core particle: scattering cross-sections in the Rayleigh regime scale with the sixth power of the particle radius (and absorption with the third power) \cite{Bohren:2004aa}. Yet, in applications such as photocatalysis that rely on energy conversion between light and hot carriers within the satellite particles, the comparatively much smaller absorption cross-section of the satellite particles is of critical importance.

Computer simulations are routinely used to predict and describe the detailed mechanisms of energy transfer between light and such nanostructures and to further improve their efficiency toward practical applications. Despite the vast array of available methods to solve Maxwell's equations in nanostructures, such as the Discrete Dipole Approximation, Surface Integral Equation, superposition $T$-matrix, Finite-Difference Time-Domain and Finite Element methods\cite{yoon_surface_2013,holler_protein-assisted_2016}, they remain very time-consuming and resource-intensive, especially in core-satellite geometries with small gaps and relatively large numbers of satellite particles. Discretisation-based methods require very fine meshes on each satellite particle to accurately predict their contribution to absorption, while mesh-free methods, such as the superposition $T$-matrix method or the closely related generalised Mie theory, require large orders of spherical wave expansions to accurately capture interparticle coupling effects.

Here, we propose and validate an alternative method based on an extension of the coupled-dipole approximation, which provides accurate results for small satellites, using the rigorous Mie theory to describe the influence of the large core particle on the satellites' optical response. Crucially, the core particle enhances the incident light due to plasmon resonances, but it also affects electromagnetic interactions between satellites due to their proximity to the metal core. Combining coupled-dipole equations with the Mie theory allows us to accurately describe these sphere-mediated interactions. The coupled-dipole approximation affords great flexibility in the number of satellites that can be considered on a standard desktop computer, ranging from a single satellite to a relatively dense coverage of hundreds of satellites around a 60\,nm core particle.

The coupled-dipole approximation is known to be limited to small particles, separated by at least a full diameter from their nearest neighbour \cite{markel_extinction_2019}. Beyond this regime, the accuracy of the results deteriorates gradually, and new spectral features can appear that are due to multipolar interactions not captured in the dipole approximation. It is not obvious \emph{a priori} whether the presence of the core particle will affect the range of validity of this approximation. We therefore performed comprehensive tests against rigorous solutions of the Maxwell equations to assess the range of validity of the method. We gave particular attention to the accurate calculation of the partial absorption in the satellites, a key physical parameter in applications such as photocatalysis.
\section{Generalised Coupled Dipole Model for core-satellite structures}
The satellite particles are described as a collection of polarisable point dipoles, where the induced dipole moment $\vecp_i=\alpha\vecE_i$ of satellite ($i$) responds to the net field $\vecE_i$ exciting it. The field $\vecE_i$ comprises the incident field $\vecEinc$, taken as a plane wave, the scattered near-field from the core sphere, $\vecEsph$, and the self-consistent field scattered by all the neighbouring satellites. The coupling between satellites can be cast in a linear system for the electric field exciting each dipole,
\begin{equation}
	\vecE_i = \vecEinc + \vecEsph + \sum_{j \neq i} \opG_{ij}\bbalpha_j\vecE_j + \sum_{j} \opS_{ij}\bbalpha_j\vecE_j,
	\label{eq:cdeshellmaintext}
\end{equation}
where \(\opG_{ij}\) is a standard dipole-dipole interaction Green's tensor in a homogeneous medium \cite{Jackson:2021aa,markel_extinction_2019}, and \(\opS_{ij}\) is a tensor of the dipole-dipole interaction mediated by the sphere \cite{auguie_electromagnetic_2019}. The latter is calculated rigorously using an extension of Mie theory for dipolar emitters. More details on the computational method are presented in Appendix \ref{AppGCDM}, and in Ref.~\cite{auguie_electromagnetic_2019}.

This Generalised Coupled-Dipole Model (GCDM) was originally developed to describe the optical properties dye molecules adsorbed on a nanoparticle \cite{auguie_electromagnetic_2019}, where each molecule was represented as a polarisable dipole with anisotropic polarisability, to account for orientation effects. Here, in contrast, we consider spherical nanoparticles for the satellites. The polarisability $\alpha$ of a small sphere of radius \(a\), dielectric function \(\varepsilon_2\) in medium with \(\varepsilon_1\) is obtained from Mie theory\cite{doyle_optical_1989,okamoto_light_1995},
\begin{equation}
    \alpha = \frac{3\ii}{2k^2} a_1,
    \label{eq:alpha_a1}
\end{equation}
where $k=2\pi\sqrt{\epsilon_1}/\lambda$ is the wavenumber in the incident medium,
\begin{equation}
    a_1 = \frac{m\psi_1(mX)\psi'_1(X)-\psi_1(X)\psi'_1(mX)}{m\psi_1(mX)\xi'_1(X)-\xi_1(X)\psi'_1(mX)},
    \label{eq:a1}
\end{equation}
\(X = ka\) is the size parameter, \(m = \sqrt{\varepsilon_2/\varepsilon_1}\) is the complex relative refractive index, and \(\psi_1(x), \xi_1(x)\) are the Riccati-Bessel functions with prime denoting derivative. Note that this polarisability prescription, obtained from rigorous Mie theory, intrinsically satisfies energy conservation, i.e. it does not require radiative correction \cite{le_ru_radiative_2013,markel_extinction_2019}. We note that Mie theory can also describe coated spheres \cite{Le-Ru:2009aa,schebarchov_simple_2013}, with only minimal changes to the model described in this work. This may be useful to account for the presence of a capping layer, or as a simplified model of nonlocal effects for very small satellites \cite{Luo:2013aa}.

From the solution of the linear system Eq.~\ref{eq:cdeshellmaintext}, we can compute the scattering and absorption properties of the core-satellite structure, as detailed in Appendix \ref{sec:cross-sections}.

In the context of photocatalytic applications \cite{Camargo:2021aa}, or photothermal applications \cite{Mackowski:1990aa}, the absorption cross-section of the particle cluster is of greater relevance than scattering or extinction; we therefore focus on absorption in this work. In practice, optical cross-sections for the whole cluster are of limited interest, as they are dominated by the response of the core particle. In comparing with experimental results, it is therefore more instructive to consider the \emph{differential} cross-sections, obtained by subtracting the optical cross-section from the bare core nanoparticle. Experimentally, this would correspond to subtracting a reference spectrum acquired from a solution of the core particles, without satellites \cite{Darby:2015aa,Stefancu:2023aa}.

The differential absorption cross-sections $\sigma_\mathrm{abs}^{\mathrm{diff}}$ presented below are defined as the difference between the total absorption of the coupled system and the absorption of the bare sphere (with no satellites),
\begin{equation}
	\sigma_\mathrm{abs}^{\mathrm{diff}} = \sigma_\mathrm{abs}^{\mathrm{sphere + dipoles}} - \sigma_\mathrm{abs}^{\mathrm{sphere}}.
	\label{eq:diffabs}
\end{equation}

With photocatalytic applications in mind, we also compute \emph{partial} absorption cross-sections, corresponding to the absorption of the satellites \emph{only}, but in the presence of the core particle. Physically, this corresponds to the absorption occurring within the satellites, which is the first step toward converting incident light into hot carriers that may contribute to chemical reactions at the satellites surfaces \cite{Herran:2022aa}. Such partial absorptions are physical quantities (unlike partial scattering cross-sections) \cite{Stout:2001aa}, but typically can only be measured indirectly, for example as a photo-chemical yield or local heat generated in photo-thermal experiments.

It is important to note that in the core-satellite system, the core sphere affects the net field seen by the dipoles, and the dipoles, in turn, affect the net field seen by the sphere. As a result, the differential absorption is generally not the same as the partial absorption in the satellites. The difference between the two can be traced in the modified absorption inside the core particle due to the presence of the surrounding dipoles. Only in the case of weak interaction (large separation, weakly-scattering dipoles) or non-absorbing core particle, do the differential and partial absorptions coincide.

\begin{figure}[!htpb]
  \centering
  \includegraphics[width=\columnwidth]{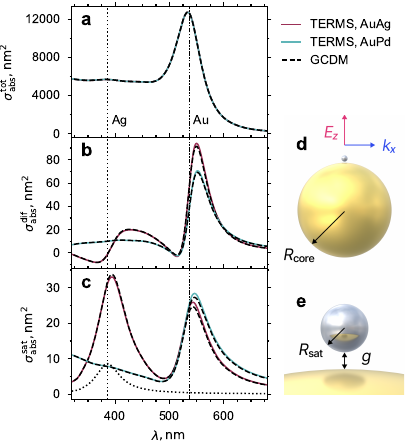}
  \caption{Absorption spectra for a single Ag or Pd satellite of radius $\Rsat = 2\,$nm, next to a spherical Au core particle of radius $\Rcore = 30\,$nm, separated by a gap $g = 1\,$nm. The particles are immersed in water. (a) Total absorption cross-section for the whole structure. (b) Differential absorption cross-section (Eq.~\ref{eq:diffabs}). (c) Partial satellite absorption cross-section. GCDM calculations (dashed lines) are compared to the rigorous $T$-matrix solution of Maxwell's equations using the TERMS program \cite{schebarchov_multiple_2022} (solid lines). The vertical dotted lines indicate the spectral position of the plasmon resonance of an isolated Au or Ag sphere in water, for reference. The absorption spectrum of an isolated Ag satellite in water is also show for comparison (dotted curve). (d) Geometry of the particle with a single Ag satellite on the Au core and the incident light propagating along the \(x\) axis with \(z\)-polarised electric field. (e) Close-up view of the satellite near the core particle.}
  \label{fig:FigA-AuAg_AuPd}
\end{figure}
\section{Application of the model and validation}
We now examine the applicability of the GCDM model to a metallic core-satellite structure with a spherical core nanoparticle surrounded by small spherical satellites. To assess the range of validity of the model, we compare the results against reference calculations obtained using the superposition $T$-matrix method, as implemented in the TERMS program \cite{schebarchov_multiple_2022}. The latter solves Maxwell's equations exactly for a collection of scatterers, but is much more demanding in computer resources than the GCDM, as illustrated in Sec.~\ref{sec:high-n} below. The benchmark results obtained with TERMS were tested for convergence and validated with a fully-independent software package, Scuff-EM, implementing the surface-integral equation \cite{Homer-Reid:2013aa}.

To simplify the presentation, we chose a representative model system consisting of a core Au sphere of radius \(\Rcore = 30\)\,nm surrounded by small satellites (1 to 4\,nm in radius). For the satellites, we considered both silver and palladium. Silver provides the strongest plasmonic response and allowed us to stress-test the method in a challenging case with strong core-satellite and satellite-satellite interactions, though it appears less directly relevant to current experimental pursuits. Palladium, in contrast, is widely used in photocatalytic applications but presents no identifiable spectral features in the visible spectrum, and we therefore chose to focus on silver in the subsequent figures to identify more easily the effects of electromagnetic interactions between particles.

The dielectric function of small metal nanoparticles is affected by the reduced mean free path of electrons compared to bulk \cite{Kreibig:2010aa}, and we therefore use a size-corrected dielectric function for the Ag satellites \cite{Yang:2015aa}. A similar correction could be applied to Pd, however, its dielectric function is poorly described by a simple Drude model, and we therefore chose to keep the bulk values from Ref.~\cite{Rakic:1998aa} for simplicity. The particles are immersed in water, described as a homogeneous non-absorbing medium of refractive index $n=1.33$.

\subsection{Single satellite}
\begin{figure*}[!htpb]
    \centering
    \includegraphics[width=\textwidth]{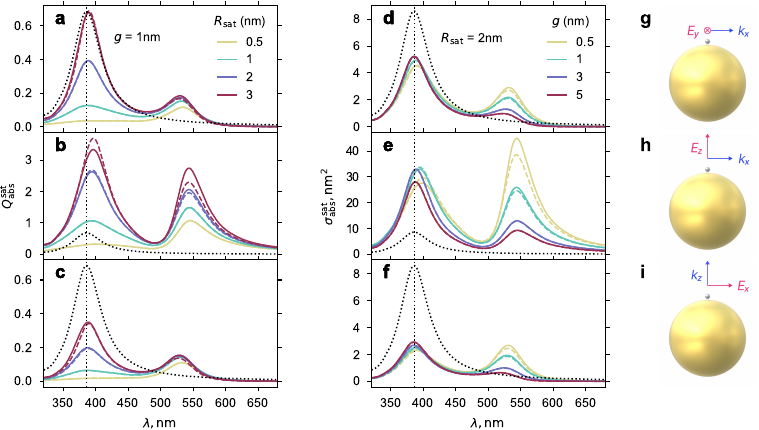}
    \caption{Effect of the satellite radius $\Rsat$ for a fixed gap of $g=1\,$nm (a--c) and the gap between the satellite and the core sphere for a fixed radius of $\Rsat=2\,$nm (d--f) on partial absorption spectra, for a Au-Ag core-satellite structure. Solid lines correspond to rigorous spectra calculated with TERMS, and dashed lines using GCDM. In panels (a--c), the partial absorption cross-section $\sigma^\mathrm{sat}_\mathrm{abs}$ is normalised by the geometrical cross-section $\sigma_\mathrm{geo} = \pi R_\mathrm{sat}^2$. The unit-less partial absorption efficiency $Q^\mathrm{sat}_\mathrm{abs} := \sigma^\mathrm{sat}_\mathrm{abs} / \sigma_\mathrm{geo}$ is used to ease the comparison between different satellite radii, as the absorption augments rapidly with the particle radius. The dotted line shows the position of the main peak of a single silver satellite without a core particle, for reference. The absorption spectrum of an isolated Ag satellite in water is also show for comparison (dotted curve, identical across all panels). Rows of the plots (a--f) correspond to different cases of the incidence on the particle illustrated in schemes (g--i).}
    \label{fig:FigC-AuAg}
\end{figure*}

The configuration depicted in Fig.~\ref{fig:FigA-AuAg_AuPd} consists of a core sphere of radius \(\Rcore = 30\)\,nm  with a single spherical satellite of radius \(\Rsat = 2\)\,nm separated from the core particle by a gap \(g = 1\)\,nm. For simplicity we modelled the optical response for a single direction of incidence, with incident electric field along the dimer axis (Fig.~\ref{fig:FigA-AuAg_AuPd} (d)). This configuration yields the strongest core-satellite interaction (see Fig.~\ref{fig:FigC-AuAg} for two orthogonal orientations).
As expected, the system with a Ag satellite differs considerably in its optical response from the Pd satellite, for both differential and partial absorption spectra (Fig.~\ref{fig:FigA-AuAg_AuPd} panels (b) and (c)). The spectrum for total absorption in panel (a) is very similar in both cases, however, since the large Au core dominates the absorption by 2 orders of magnitude.

The dominant feature around \(\lambda \approx 550\)\,nm in the differential absorption spectra for both Ag and Pd satellites reflects the shift of the Au plasmon resonance due to the presence of the satellite (a red-shift in both cases here). The differential spectrum for Ag has a pronounced bisignate, derivative-like feature around 390\,nm. The zero-crossing point occurs at the resonance position of a 2\,nm Ag sphere immersed in water. We can interpret this feature by considering the response of coated Au core particle, where the coating layer here consists of a very low concentration of satellites (a single one) \cite{Darby:2015aa,Tang:2021aa}. The effective dielectric function of the coating at low satellite concentration is simply proportional to the satellite polarisability $\alpha$, with its real part crossing zero at the resonance position. The effective coating therefore presents a lower (resp. higher) refractive index compared to the embedding medium seen by the core particle on either side of the Ag resonance, which leads to a lower (resp. higher) absorption compared to the bare core particle in water.

The partial satellite absorption spectra (panel c) differ markedly from the differential spectra (panel b). With the Ag satellite, the plasmon resonance of Ag is clearly visible at \(\lambda \approx 400\)\,nm, while Pd presents a relatively featureless absorption in this region. For both Ag and Pd satellites, a strong absorption peak is observed \emph{in the satellite} in the spectral range of the core Au plasmon resonance ($550\,$nm). We attribute this feature to the enhancement of the satellite's internal field via the plasmon resonance of the neighbouring core particle. Overall, differential absorption demonstrates a redistribution of energy in the whole cluster with respect to the bare core, while partial absorption provide more detailed information on the absorption occurring in specific parts of the structure.

\begin{figure}[!htpb]
  \centering
  \includegraphics[width=\columnwidth]{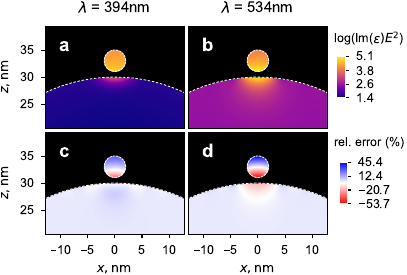}
  \caption{Near-field intensity maps at resonance for the absorption by a Au core and a Ag satellite for two wavelengths: \(\lambda = \)394\,nm (a) and \(\lambda = \)534\,nm (b). The calculations were performed with TERMS, using a maximum multipolar order of 40. The bottom panels (c,d) show the corresponding relative error resulting from using a dipole approximation to describe the satellite's optical response. This was obtained by simulating the cluster with TERMS and setting the $T$-matrix coefficients describing the satellite particle, but not the core particle, to zero, for all but the electric dipole coefficient. The error is obtained by subtracting the converged solution displayed above. Note the linear scale in panels (c--d).
}
  \label{fig:near-field}
\end{figure}
To better understand the distribution of electromagnetic energy in the Au-Ag structure, we used the TERMS program to compute near-field intensity maps at the two resonance wavelengths (Fig.~\ref{fig:near-field}). Specifically, we display the spatial distribution of absorption inside the cluster depicted in Fig.\ref{fig:FigA-AuAg_AuPd} (d), noting that absorption of electromagnetic energy is proportional to $\mathrm{Im}(\varepsilon)|E|^2$, where $\varepsilon$ is the relative dielectric function at the relevant frequency, and $|E|^2$ the electric field intensity. Despite the very small size of the satellite particle, its internal field presents a strong gradient in the $z$ direction (radially from the core particle). This can be attributed to the highly inhomogeneous field surrounding the core particle. From this observation, it is remarkable that the dipolar approximation, on which the GCDM is based, is able to predict the average satellite's absorption with good accuracy in Fig.~\ref{fig:FigA-AuAg_AuPd}, even though it considers the satellite as a point dipole. The bottom panels of Fig.~\ref{fig:near-field} offer an explanation for this fortuitous accuracy obtained in the average satellite response. The simulations in panels (c--d) were performed with TERMS by replacing the satellite particle with a pure electric dipole \cite{schebarchov_multiple_2022}, while keeping a full multipolar response for the core particle (a maximum multipolar order of 40 was used throughout). This truncated multipolar response for the satellite response closely matches the GCDM. The error in this approximation is obtained by comparison with the fully-converged numerical results, where both satellite and core particles use a multipolar truncation order of 40. In the dipole approximation, the field inside the satellite should be constant, while the full numerical solution reveals a strong vertical gradient. To first order, for sufficiently small satellite particles, the internal field is linearly underestimated at the bottom, and overestimated at the top of the satellite, and the average absorption over the whole volume is predicted accurately.

The near-field maps also highlight the effect of the satellite on the core particle, which is required to account for the difference between partial and differential absorption spectra: the presence of the satellite affects the internal field distribution inside the core particle. Given how localised this effect is, a dipole-only approximation for the core particle would certainly fail, highlighting the necessity of considering the full exact Mie solution for the core response, up to sufficiently high order. Numerically, we found that a maximum order of 40 to 50 is sufficient for convergence in this study. For smaller core-satellite gap distances, much higher values can be required \cite{auguie_electromagnetic_2019}. This is not problematic for the GCDM, where very high multipolar orders can be used \cite{Majic:2020aa,auguie_electromagnetic_2019}, but we note that the superposition $T$-matrix method, as implemented in TERMS, starts to suffer numerical accuracy problems beyond 40 or 50 \cite{Schebarchov:2019aa}.

A key requirement of the coupled-dipole approximation, and by extension of the GCDM, is that the scatterer should be small enough to be described as a point dipole, and that neighbouring scatterers should be sufficiently separated to not induce a strong multipolar response beyond the dipole approximation. To assess the range of applicability of the GCDM, we therefore varied the satellite radius, and its distance to the core particle surface (Fig.~\ref{fig:FigC-AuAg}). The GCDM predictions (dashed lines) are compared to rigorous solutions obtained using TERMS (solid lines). As expected the GCDM is more accurate for smaller sizes of satellites and large gaps (weaker core-satellite interaction). The accuracy of the GCDM remains very good over a good range of parameters relevant to experiments. The discrepancy between GCDM and TERMS  becomes significant for a satellite radius \(\Rsat \ge 3\)\,nm (at a gap of 1\,nm) or for a gap $g\le 0.5$\,nm (at a radius of 2\,nm). The same conclusions can be drawn regardless of the direction of incidence and polarisation, although naturally, the absolute magnitude of the partial absorption varies strongly with polarisation.
\subsection{Dimer of satellites}
\begin{figure}[!htpb]
    \centering
    \includegraphics[width=\columnwidth]{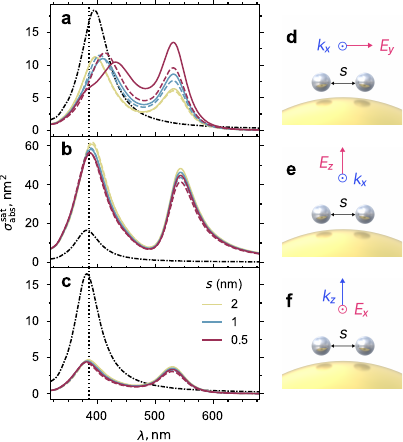}
    \caption{Partial absorption spectra for two satellites with varying separation \(s\). Solid lines correspond to spectra calculated with TERMS and dashed lines with GCDM. The dotted vertical line shows the position of the main peak of a single silver satellite without a core particle. The dash-dotted spectrum shown for reference corresponds to the two silver satellites with no core Au particle present. Panels (a--c) correspond to different orientations of the incident light, illustrated in the corresponding schemes (d--f).}
    \label{fig:FigD-AuAg}
\end{figure}

To understand the effect of multiple satellites on the response of a core-satellite nanostructure we start with the simplest configuration of two satellites separated by a distance \(s\) (Fig.~\ref{fig:FigD-AuAg}).
Both satellites are located symmetrically above the core sphere with a gap \(g\). The GCDM model should be most suitable for well separated satellites when their interaction can be described as dipole-dipole coupling. Below a certain separation between satellites, higher order multipolar interactions may become important, leading to a gradual decrease of accuracy, and the possible emergence of new spectral features not captured by the coupled-dipole approximation. As shown in Fig.~\ref{fig:FigD-AuAg}, for a pair of identical satellites with radius 2\,nm and a gap $g=1\,$nm to the core surface, discrepancies become important for \(s\lessapprox 1\)\,nm. The non-dipolar interactions appear to be particularly strong when the incident electric field is polarised along the axis of the satellite dimer (\(y-\)axis), which is also observed in the standard coupled-dipole model (no core particle present).

Having established a minimum "safe" separation of two satellites allows us to formulate a "rule of thumb" for the applicability of the GCDM in this configuration. Fig.~\ref{fig:FigD-AuAg} suggests that accurate results may be obtained when the distance between satellites is greater than $2$\,nm.
\subsection{High satellite coverage}
\label{sec:high-n}
\begin{figure}[!htpb]
    \includegraphics[width=\columnwidth]{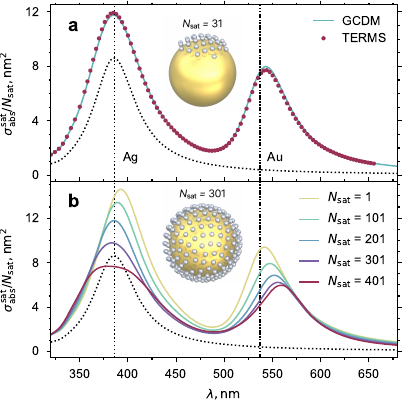}
    \caption{Influence of the number of satellites \(N_\mathrm{sat}\). (a) Validation of the GCDM model against TERMS for the partially covered core particle with a dense coverage of 31 satellites equivalent to a full coverage of standard core-satellite structure with 301 satellites (structure shown in inset).
    (b) GCDM calculations for a fully-covered spherical core, with varying numbers of satellites. Spectra were calculated using numerical orientation averaging \cite{Fazel-Najafabadi:2022aa} for the incident light. Correspondences of the number of satellites and their minimal inter-satellite separation are as follows: \(N_\mathrm{sat} = 101, s \approx 6.2\)\,nm; \(N_\mathrm{sat} = 201, s \approx 3.2\)\,nm; \(N_\mathrm{sat} = 301, s \approx 1.9\)\,nm; \(N_\mathrm{sat} = 401, s \approx 1.1\)\,nm.}
    \label{fig:FigF-AuAg}
\end{figure}
With a core particle surrounded by many satellites, we may ask whether the dipole-dipole interactions are dominated by nearest-neighbour effects, or perhaps involve collective interactions between more satellites that could invalidate the above "rule of thumb". We therefore considered the effect of a relatively dense coverage of the core particle, and assessed the validity of the model against TERMS. With multiple satellites, many different configurations around a spherical core could be considered, ranging from ordered to disordered coverage \cite{Auguie:2018aa}. For simplicity and clarity, we focused on a relatively ordered coverage, where the nearest-neighbour distance is very uniform. Specifically, we used the Fibonacci lattice described in Ref.~\cite{Hardin2016ACO} which places an odd number of points on the surface of a sphere. For our model system with \(\Rcore = 30\)\,nm, \(\Rsat = 2\)\,nm, \(g = 1\)\,nm, a number of satellites \(N_\mathrm{sat} = 301\) yields a minimum separation between pairs of satellites of \(s_\mathrm{min} \approx 1.9\)\,nm. The strict validation of the GCDM model against TERMS for this number of satellites requires unrealistically large computational resources for the TERMS calculations. For a more pragmatic validation, we therefore compared results for a spherical cap of 31 satellites, with equal satellite density as a core sphere fully covered by 301 satellites. The structure with 31 satellites is depicted in the inset of Fig.~\ref{fig:FigF-AuAg} (a) alongside the comparison of partial absorption spectra per satellite. GCDM results are in good agreement with TERMS at this satellite density, demonstrating that the rule of thumb of inter-satellite separation described above for 2 satellites extends to multiple satellite configurations.

With the GCDM results validated in this multi-satellite configuration, we now explore the effect of satellite concentration on the partial absorption spectra (Fig.~\ref{fig:FigF-AuAg} (b)). For clarity, the cross-sections are normalised by the number of satellites covering the spherical core. Note that the spectrum for \(N_\mathrm{sat} = 401\) is expected to be slightly incorrect, since the minimum separation \(s \approx 1.1\)\,nm is likely too small for the coupled-dipole approximation to be accurate. Overall, the partial absorption spectrum per satellite does not change significantly up to the fairly high coverage considered here. For \(N_\mathrm{sat}>100\), dipole-dipole interactions result in a small decrease in the absorption per satellite at both resonances. The resonances are also slightly shifted and broadened as coverage increases, to the red for the gold-core resonance and to the blue for the satellite resonance \cite{auguie_electromagnetic_2019}. In this weak satellite-satellite interaction regime core-satellite nanostructures could achieve a consistent optical response without requiring a very uniform spacing between satellites \cite{rossner_planet-satellite_2020}. These considerations can inform the design of core-satellite nanostructures for specific applications, and help with their electromagnetic modelling.

The primary advantage of the GCDM over the more rigorous $T$-matrix method lies in its higher computational speed and smaller memory footprint for multiple satellites. For the simulations presented in this paper we used a high-end personal computer with the following specifications: 12 cores Intel(R) Xeon(R) CPU E5-1650 v3 \@ 3.50GHz, 64GB System Memory DDR4, running under Ubuntu 18.04.3. Although the computation time for the TERMS calculation of a single satellite at a single wavelength is relatively small (9.8\,s), it increases significantly with the number of satellites. For the structure with 31 satellites illustrated in Fig.~\ref{fig:FigF-AuAg} (a) the computational time was 50 minutes for a single wavelength and required 57\,GB of RAM. Routine calculations therefore rapidly become impractical on standard computers, when both a high multipolar order and a high number of particles are required. We note that this problem is partly specific to our $T$-matrix implementation, and a more specialised code could in principle be designed to lower the memory footprint. In contrast, our GCDM implementation in the Matlab environment computes the same 31 satellite structure in 0.027\,s per wavelength, using 2.43\,GB of memory.

\section{Conclusions}
\label{sec:conclusions}

Our proposed generalised coupled-dipole model (GCDM) offers an efficient means of calculating the optical response of core-satellite nanostructures with a large number of small satellites, which is otherwise impractical using more rigorous simulation methods. The model allows detailed and accurate analysis of such systems with the ability to calculate both partial and differential absorption spectra, offering complementary insights into the dissipation of energy inside the nanostructure. We successfully applied this model to a computationally challenging system consisting of a core gold sphere surrounded by small silver satellites and validated our results against the rigorous superposition $T$-matrix method. We established the range of applicability of the model by varying all key parameters: satellite radius, distance to core, incidence direction, and satellite coverage. The method offers considerable benefits in its computational speed and memory footprint when compared to the $T$-matrix method; it is also a much simpler method to implement in computer code. The GCDM present a powerful and efficient approach for evaluating the optical response of core-satellite nanostructures, with a particular focus on absorption characteristics that are relevant to photocatalysis and photothermal experiments. These advancements can help improve the design and understanding of complex nanoparticle assemblies, opening many possibilities for future research and applications in nanotechnology.
\section*{Conflicts of interest}
There are no conflicts to declare.
\section*{Acknowledgements}
The authors thank the Royal Society Te Ap\=arangi for support through Marsden grants MFP-VUW2204 and MFP-VUW2118.

%
\appendix
\label{AppGCDM}
\section{Generalised Coupled-Dipole Model}
\label{sec:gcdm}
We summarise below for completeness the equations of the GCDM introduced in Ref.~\cite{auguie_electromagnetic_2019}.
\subsection{Coupled-dipole equations}
\label{sec:cda}
We first introduce our notations for the standard equations for the coupled-dipole model in a homogeneous medium. We consider a collection of \(N\) point polarisable dipoles located at positions \(\vecr_i\) embedded in an infinite homogeneous medium characterised by a dielectric function \(\varepsilon_1=n^2\). The response of a point dipole to the electric field is linear and defined by the polarisability $\alpha$,
\begin{equation}
	\vecp_i = \alpha \vecE(\vecr_i).
\end{equation}
This induced dipole \(\vecp_i\) generates in turn an electric field at a
general point \(\vecr\) that also depends linearly on the dipole moment,
\begin{equation}
	\vecE_{\vecp_i}(\vecr) =  \opG(\vecr_i,\vecr)\, \vecp_i,
	\label{eq:EGp}
\end{equation}
where the Green's tensor \(\opG\) characterises the electric field at \(\vecr\) created by unit point dipoles along $x,y,z$ placed at the location \(\vecr_i\).

The net electric field acting on each dipole consists of the incident field \(\vecEinc\), taken here as a plane wave, and the field scattered by the neighbouring dipoles. This results in a system of coupled-dipole equations for the self-consistent fields \(\vecE_{i}\),
\begin{equation}
	\vecE_i =  \vecEinc(\vecr_i) + \sum_{j \neq i} \opG_{ij}\bbalpha\vecE_j.
	\label{eq:cde_E}
\end{equation}
where \(\opG_{ij} := \opG(\vecr_i,\vecr_j)\) is the Green's tensor coupling dipoles \(j\) and \(i\) in the infinite surrounding medium \cite{Jackson:2021aa},
\begin{equation}
	\opG_{ij} =  \beta^{-1}\frac{\expikr_{ij}}{r_{ij}}\Bigg\{k_1^2\left[\opI - \vecrh_{ij}\otimes\vecrh_{ij} \right] - \left(\frac{1}{r_{ij}^2} - \frac{\ii k_1}{r_{ij}}\right) \left[\opI - 3\vecrh_{ij}\otimes\vecrh_{ij}\right] \Bigg\}
	\label{eq:Gij}
\end{equation}
\(k_1 = 2\pi \sqrt{\varepsilon_1}/\lambda\) is the wave vector in the embedding medium, and we introduce a prefactor \(\beta = 4\pi\varepsilon_0\varepsilon_1\) which may be simplified throughout in practice by defining suitably-normalised quantities.

After solving the system of coupled dipole equations Eq.~\ref{eq:cde_E} for the electric fields \(\vecE_i\) and then calculating dipole moments \(\vecp_i\) the absorption cross-section of the dipoles is given by \cite{le_ru_principles_2008}:

\begin{equation}
	\Cabs = \frac{ 4\pi\beta^{-1} k_1} {|E_0|^2} \sum_i \left(\Im\left[  \vecp_i \cdot \vecE_i^* \right] - \frac{2\beta}{3}k_1^3 |\vecp_i|^2 \right),
	\label{eq:Cabs}
\end{equation}
where \(E_0\) is the amplitude of the incident electric field.
\subsection{Generalised coupled-dipole equations}
\label{sec:GCDM}
We now consider a modified system consisting of the same collection of point dipoles augmented by a homogeneous sphere to describe the core particle. The resulting field acting on the dipole \(\vecp_i = \bbalpha_i\vecE_i\) now also has a contribution of the incident field scattered by the sphere, \(\vecEsph\), and the field produced by the interaction of the dipoles with the sphere,
\begin{equation}
	\vecE_i = \vecEinc + \vecEsph + \sum_{j \neq i} \opG_{ij}\bbalpha_j\vecE_j + \sum_{j} \opS_{ij}\bbalpha_j\vecE_j,
	\label{eq:cdeshell}
\end{equation}
where \(\opG_{ij}\) is the previous Green's tensor from Eq.~\ref{eq:Gij}, while \(\opS_{ij}\) is a tensor characterising the dipole-dipole interaction mediated by the sphere. We calculate \(\vecEsph\) and \(\opS_{ij}\) numerically using Mie theory (see Ref.~\cite{le_ru_principles_2008} and \cite{auguie_electromagnetic_2019}). We note that \(\opS_{ii}\) is not zero; it represents a self-interaction of dipole \(i\) caused by the presence of the sphere (similar to a "reflected field" acting back on the dipole itself).

Grouping the fields \(\vecE_i, \vecEinc\), and \(\vecEsph\) in these \(3N\) equations we can write a matrix form of the coupled-dipole equations,
\begin{equation}
	\opA \vecE = \vecEinc + \vecEsph,
	\label{eq:AE}
\end{equation}
where \(\opA\) is a full interaction matrix that combines the 3\(\times\)3 Green's tensors \(\opG_{ij}\) and \(\opS_{ij}\),
\begin{equation}
	\opA_{ij} = \begin{cases}
		\opI_3 - \opS_{ii}\bbalpha_i  & i=j,\\
		\opG_{ij}\bbalpha_j - \opS_{ij}\bbalpha_j  & i\neq j.
	\end{cases}
	\label{eq:Aij}
\end{equation}
Numerical solutions of Eq.~\ref{eq:AE} for \(\vecE_i\) allow to obtain the dipole moments \(\vecp_i\) as in the standard coupled-dipole theory.

We now describe the additional terms introduced in the generalised coupled-dipole system (Eq.~\ref{eq:cdeshell}) by the introduction of the sphere, as well as the calculation of far-field cross-sections for the combined system in the framework of generalised Mie theory.

\subsection{Exciting field due to the sphere.}

The net incident field on the collection of dipoles (source term in the linear system of Eq.~\ref{eq:AE}) is augmented by the contribution $\vecEsph$. For a given external incident field, such as a plane wave propagating along an arbitrary direction, we use the standard Mie theory\cite{Bohren:2004aa,Le-Ru:2009aa} to compute the electric field scattered by the bare sphere at any point in space, and in particular at the location of each dipole.

\subsection{Coupling mediated by the sphere.}

The Green's tensor $\opS_{ij}$ expresses the field created at location $\vecr_j$ by a unit dipole at $\vecr_i$ due to scattering by the sphere, with the dipole placed in three orthogonal orientations for each column of $\opS_{ij}$. We calculate this $3\times 3$ matrix using the generalised Mie theory, where the field of the exciting dipole $\vecp_i$ is decomposed in a basis of vector spherical wavefunctions (VSWFs) centred on the sphere\cite{MishchenkoTL02,Le-Ru:2009aa},
\begin{align}
\vecE^\text{i,\sc dip} &= \sum_{n=1}^{\infty}\sum_{m=-n}^{n} a^\text{i,\sc dip}_{mn} \vecrgM(k_1, \vecr) + b^\text{i,\sc dip}_{mn} \vecrgN(k_1, \vecr),\qquad r \leq r_i, \label{Edips}\\
\vecE^\text{i,\sc dip} &= \sum_{n=1}^{\infty}\sum_{m=-n}^{n} e^\text{i,\sc dip}_{mn} \vecM(k_1, \vecr) + f^\text{i,\sc dip}_{mn} \vecN(k_1, \vecr),\qquad r \geq r_i.
\label{eq:ef}
\end{align}
$(\vecN, \vecM), (\vecrgN, \vecrgM)$ are the regular and irregular VSWFs, respectively, $a^\text{i,\sc dip}_{mn}$, $b^\text{i,\sc dip}_{mn}$, $e^\text{i,\sc dip}_{mn}$, $f^\text{i,\sc dip}_{mn}$ are expansion coefficients with known analytical expressions in terms of $\vecp_i$ (given in Sec.~\ref{sec:appexpansion}). The second expansion in terms of irregular waves will be used in the calculation of cross-sections.

The field scattered by the sphere from this dipolar excitation follows from Mie theory,
\begin{equation}
\vecEsca = \sum_{n=1}^{\infty}\sum_{m=-n}^{n} p_{mn} \vecM(k_1, \vecr) + q_{mn} \vecN(k_1, \vecr)
\label{Esca}
\end{equation}
with
\begin{align}
p^\text{i,\sc dip}_{mn} &= \Gamma_{n} a_{mn}^\text{i,\sc dip}, \\
q^\text{i,\sc dip}_{mn} &= \Delta_{n} b_{mn}^\text{i,\sc dip},
\end{align}
where $\Gamma_{n}$ and $\Delta_{n}$ are the standard electric and magnetic multipolar Mie susceptibilities\cite{Bohren:2004aa,Le-Ru:2009aa}.

\begin{align}
\Gamma_{n} = \frac{m\psi_n(X)\psi'_n(mX)-\psi_n(mX)\psi'_n(X)}{\psi_n(mX)\xi'_n(X)-m\xi_n(X)\psi'_n(mX)}, \\
\Delta_{n} = \frac{\psi_n(X)\psi'_n(mX)-m\psi_n(mX)\psi'_n(X)}{m\psi_n(mX)\xi'_n(X)-\xi_n(X)\psi'_n(mX)}
\end{align}
(the coefficient $a_1$ appearing in Eq.~\ref{eq:a1} is equal to $\Delta_1$ in these more general notations).


%
\subsection{Far-field cross-sections.}
\label{sec:cross-sections}

A formal expansion similar to Eq.~\ref{Edips} is used for the plane wave illumination at arbitrary incidence (with different, known coefficients\cite{MishchenkoTL02}) corresponding to the incident field,
\begin{equation}
\vecEinc = \sum_{n=1}^{\infty}\sum_{m=-n}^{n} a^\text{\sc pw}_{mn} \vecrgM(k_1, \vecr) + b^\text{\sc pw}_{mn} \vecrgN(k_1, \vecr).
\end{equation}
The sum of both incident and dipole fields forms the net exciting field for the sphere,
\begin{equation}
\begin{split}
\vecEexc & = \vecEdips + \vecEinc \\
&= \sum_{n=1}^{\infty}\sum_{m=-n}^{n}  a_{mn}^\text{\sc exc}\vecrgM(k_1, \vecr) + b_{mn}^\text{\sc exc} \vecrgN(k_1, \vecr)
\end{split}
\label{Eexc}
\end{equation}
where we simply sum all dipole coefficients to those of the incident plane wave,
\begin{equation}
\begin{aligned}
a_{mn}^\text{\sc exc} &= a^\text{\sc pw}_{mn} + \sum_i a^\text{i,\sc dip}_{mn},\quad \text{and}\\
b_{mn}^\text{\sc exc} &= b^\text{\sc pw}_{mn} + \sum_i b^\text{i,\sc dip}_{mn}.
\end{aligned}
\label{eq:abexc}
\end{equation}
The coefficients for the field scattered by the sphere from the combined excitation follow from Mie theory,
\begin{align}
p_{mn} &= \Gamma_{n} a_{mn}^\text{\sc exc}, \\
q_{mn} &= \Delta_{n} b_{mn}^\text{\sc exc}.
\end{align}
The total scattering cross-section is obtained by summing the total field scattered by the sphere, and that directly radiated by the dipoles (Eq.~\ref{eq:ef}), both expressed in a basis of irregular VSWFs.
\begin{equation}
\Csca = \frac{1}{k_1^2}\sum_{n=1}^{\infty}\sum_{m=-n}^{n} |e^\text{\sc dip}_{mn}+p^\text{\sc sph}_{mn}|^2 + |f^\text{\sc dip}_{mn}+q^\text{\sc sph}_{mn}|^2
\label{sca}
\end{equation}

For the extinction cross-section we invoke the optical theorem, where the incident field is a plane wave excitation along a specific direction, and the scattered field is given by the superposition of the dipole sources and the total field scattered by the sphere, which results in:
\begin{equation}
\Cext = \frac{-1}{k_1^2}\sum_{n=1}^{\infty}\sum_{m=-n}^{n} \Re\left(a_{mn}^\text{\sc pw}p^*_{mn}+ a_{mn}^\text{\sc pw}e^\text{*\sc dip}_{mn}\right) + \Re\left(b_{mn}^\text{\sc pw}q^*_{mn}+ b_{mn}^\text{\sc pw}f^\text{*\sc dip}_{mn}\right).
\label{ext}
\end{equation}

The absorption cross-section is then deduced as
\begin{equation}
\Cabs = \Cext - \Csca.
\label{abs}
\end{equation}

\subsection{Expansion coefficients for plane wave and dipole}
\label{sec:appexpansion}

Our definitions for the vector spherical wavefunctions follow Mishchenko et al.~\cite{MishchenkoTL02}; we summarise below for convenience the expansion coefficients for a plane wave illumination\cite{MishchenkoTL02} and a dipole source\cite{Le-Ru:2009aa}, with reference to the equation number in the original source,

\begin{align}
a^\text{\sc pw}_{mn}  &= 4\pi (-1)^m i^n r_{n} e^{-im\varphi} \veceh \cdot \vecCmn^*(\theta)\qquad \text{[C.57] ~in~Ref.~\citenum{MishchenkoTL02} }\\
b^\text{\sc pw}_{mn}  &= 4\pi (-1)^m i^{n-1} r_{n} e^{-im\varphi} \veceh \cdot \vecDmn^*(\theta)\\[1em]
a^\text{\sc dip}_{mn} &= E_{p0} (-1)^m \vecph \cdot \vecM_{-mn}(r,\theta,\phi)  \qquad \text{[H.84]~in~Ref.~\citenum{Le-Ru:2009aa}} \\
b^\text{\sc dip}_{mn} &= E_{p0} (-1)^m \vecph \cdot \vecN_{-mn}(r,\theta,\phi) \\[1em]
e^\text{\sc dip}_{mn} &= E_{p0} (-1)^m \vecph \cdot \vecrgM_{-mn}(r,\theta,\phi) \qquad \text{[H.86]~in~Ref.~\citenum{Le-Ru:2009aa} }\\
f^\text{\sc dip}_{mn} &= E_{p0} (-1)^m \vecph \cdot \vecrgN_{-mn}(r,\theta,\phi)
\end{align}
where we defined
\begin{equation}
r_{n} = \sqrt{\dfrac{2n+1}{4\pi n(n+1)}},
\end{equation}

and \(\vecph\) and \(\veceh\) are unit vectors in the direction of \(\vecp\)
and \(\vecE\), respectively. The prefactors for both fields are

\begin{align}
E_{0} & = 1, \quad \text{(we assume a unit incident field throughout)}\\
E_{p0} & = \frac{ik_1^3 p}{\epsnot\eps} = 4\pi ik_1^3 \,\beta^{-1} \,p , \qquad  \text{[H.83]~in~Ref.~\citenum{Le-Ru:2009aa} }
\end{align}

\bibliography{references} 
\bibliographystyle{rsc} 

\end{document}